\documentclass{aa}  
\usepackage{graphicx}
\usepackage{txfonts}
\usepackage{natbib}
\usepackage{multirow}

\bibpunct{(}{)}{;}{a}{}{,}

\begin{document}
   \title{The nearby QSO host I~Zw~1: The stellar disk and adjacent objects\thanks{Based on observations collected at the Very Large Telescope (UT1) of the European Southern Observatory, Paranal (Chile), under service mode project 67.B-0009. Complementary data are based on observations made with the 3.6m telescope of the European Southern Observatory, La Silla (Chile), and with the NASA/ESA Hubble Space Telescope, obtained from the ESO/ST-ECF Science Archive Facility (joint collaboration of the European Southern Observatory and the Space Telescope - European Coordinating Facility)}}

   \author{J. Scharw\"achter\inst{1} \and A. Eckart\inst{2} \and S. Pfalzner\inst{2} \and I. Saviane\inst{3} \and J. Zuther\inst{2}
}

\offprints{J. Scharw\"achter}

\institute{European Southern Observatory, Casilla 19001, Santiago 19, Chile\\
\email{jscharwa@eso.org}
    \and
        I. Physikalisches Institut, Universit\"at zu K\"oln, Z\"ulpicher Str. 77, 50937 K\"oln, Germany
    \and
       European Southern Observatory, Casilla 19001, Santiago 19, Chile}

\date{Received (date); accepted (date)}

% \abstract{}{}{}{}{} 
% 5 {} token are mandatory
 
\abstract
   {The relation between tidal interactions, starbursts, 
and the onset and/or fueling
of active galactic nuclei (AGN) is a matter of debate.
I~Zw~1 is considered as the prototypical 
narrow-line Seyfert~1 galaxy (NLS1) and as one of the closest
quasi-stellar objects (QSOs). With a clear spiral host and a small
companion galaxy to the west, I~Zw~1 is a possible example
of minor-merger-related nuclear activity.}
   {This study investigates possible signs of a relation between
   merger process, star formation activity, and AGN properties in
   the case of I~Zw~1. }
   {The morphology of I~Zw~1 and nearby sources is investigated
via high-resolution NIR images.
Color trends in the host galaxy of I~Zw~1 are discussed by means 
of optical-to-NIR color composites.
  Long-slit spectra of
the QSO nucleus of I~Zw~1 and of the two nearby sources to the
north and the west of the I~Zw~1 disk are analyzed.}
   {The data support the scenario of a tidal interaction between 
I~Zw~1 and the small companion galaxy to the west. 
A concentration of blue color in the western part 
of the I~Zw~1 host galaxy might be the
manifestation of merger-induced star formation activity. 
Previous findings that the likely companion has an old evolved stellar
population are substantiated by the new data. An extension to the 
west of the putative companion emerges as a separate source. 
The source to the north of the I~Zw~1 disk is reconfirmed
as a late-type foreground star. Lines in the
nuclear $K$-band
spectrum of I~Zw~1 are discussed in comparison to 
data prior to this article and line fluxes are reported.}
   {}


   \maketitle

%
%________________________________________________________________
\section{Introduction\label{sec:intro}}

It is widely accepted that nuclear activity in galaxies is
caused by mass accretion onto a supermassive black hole 
\citep[see e.g. review by][]{2004ASPC..311...49U}. Furthermore,
it is known that AGN are often connected to 
star formation activity \citep[e.g.][]{2003MNRAS.346.1055K}.
However, the importance of tidal interactions for the
onset and/or fueling of AGN and starbursts is still unclear
\citep[e.g.][]{2003MNRAS.340.1095D, 2005ApJ...627L..97G, 
2006ApJ...651...93K}. 

Regarding the comparable bolometric luminosities
and local space densities of ULIRGs and local QSOs, 
\citet{1988ApJ...325...74S} proposed an evolutionary link
between the two classes of objects. In this scenario,
galaxy mergers are assumed to cause the gas inflow which triggers
an intense starburst and a dust-enshrouded
AGN. At this stage, the object would appear as 
ULIRG. When winds and feedback blow away the dust from the central
region, the object goes through an optically bright QSO-phase. 
Such an evolutionary scenario is supported by recent numerical
simulations which include supermassive black holes and feedback
\citep[e.g.][]{2005Natur.433..604D,2005ApJ...630..705H,2006ApJS..163....1H}.
Observations show that 
virtually all ULIRGs are found in major merger systems
\citep[e.g.][]{2002ApJS..138....1B, 2002ApJS..143..315V}. But 
on the side of QSOs, the importance of mergers is far less conclusive.
In fact, there is increasing evidence that the results
might depend on the selection of the QSO sample.
A comparison of ULIRGs with
QSOs from the radio-loud/radio-quiet sample by 
\citet{2003MNRAS.340.1095D}, which are mainly located in
massive elliptical host galaxies, shows that both populations
occupy different parts of the fundamental plane 
\citep{2002ApJ...580...73T}.
Palomar Green QSOs, instead, are discussed as putative
candidates for a post-ULIRG stage 
\citep[e.g.][]{2001AJ....122.2791S,
2003A&A...402...87H,2007ApJ...657..102D}. 
An exception, however,
is the prominent fraction of 
Palomar Green QSOs with clear spiral host galaxies. 
According to today's consensus from numerical simulations,
major mergers in the local Universe typically result
in elliptical merger remnants 
\citep[e.g.][]{1992ApJ...400..460H, 2003ApJ...597..893N, 
2005MNRAS.357..753G, 2006MNRAS.369..625N}. 
Spiral hosts are very unlikely to have undergone a recent
major merger.

Nuclear activity in spiral host galaxies might instead be
related to minor mergers with low-mass companion galaxies.
Numerical
simulations show that minor mergers can cause large amplitude
spirals and radial gas inflow in the primary galaxy
\citep{1995ApJ...448...41H}. \citet{1995ApJ...448...41H}
report that this gas could
be available for the fueling of a nuclear
starburst and AGN activity but that the gas
inflow rates during the early stages of the merger
are low.
Consequently, central activity might only be induced in the
late stages of the merger, i.e. when the satellite galaxy has
reached the central few kiloparsecs of the primary galaxy. 
The different observational studies about a possible excess
of companions around Seyfert galaxies yield 
extremely contradictory 
results \citep[e.g.][]{1984AJ.....89..966D,1995AJ....109.1546R,
1995A&A...293..683L, 1998ApJ...496...93D,1999ApJ...513L.111D}.
Rather than triggering nuclear activity, minor mergers may
instead increase the luminosity of a pre-existing AGN to
QSO-levels \citep{2000ApJ...536L..73C}. 
Several observables of low-luminosity 
Seyfert galaxies 
can also be reproduced by numerical simulations
of a merger-independent model \citep{2006ApJS..166....1H}. 
In this model, the fueling
process is based on the 
stochastic accretion of
cold gas.

\begin{figure}
        \centering\resizebox{\hsize}{!}{\includegraphics{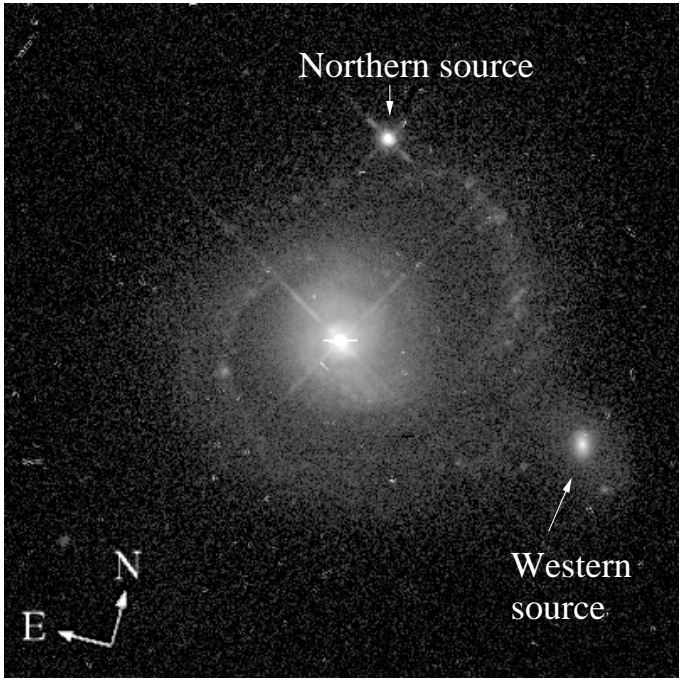}}
        \caption{HST image of I~Zw~1 based on observations with the WFPC2 camera in the F814W filter, obtained from the ESO/ST-ECF Science Archive Facility. For this plot, the image has been partially cleaned from cosmic rays and processed with an unsharp masking technique.}
        \label{fig:zwgeneral}
\end{figure}

In view of the above controversies,
I~Zw~1 is an interesting example for a 
detailed observational case study.
I~Zw~1 is classified as a NLS1 as well as an infrared-excess
Palomar Green QSO and a possible candidate for an ongoing
minor merger. With  
$M_B = -22.62$ \citep{2001ApJ...555..719C}, 
the nuclear blue magnitude of I~Zw~1
just meets the 
QSO criterion 
$M_B < -22.1$ \citep{1983ApJ...269..352S}\footnote{Assuming $H_0=75\ \mathrm{km\ s^{-1}\ Mpc^{-1}}$
and $q_0=0.5$.}.
At a redshift of
only $z=0.0611$ \citep{1985AJ.....90.1642C}, I~Zw~1 is therefore
considered as one of the closest QSOs.
As illustrated by the Hubble Space Telescope (HST) image 
in Fig.~\ref{fig:zwgeneral},
the spiral host galaxy of I~Zw~1 is clearly visible.
The host of I~Zw~1 shows direct observational 
evidence for young stellar populations
and ongoing nuclear star formation activity
\citep[e.g.][]{1990AJ.....99...37H, 1994ApJ...424..627E, 1998ApJ...500..147S,2001ApJ...555..719C,
2003A&A...405..959S}. 
The nuclear starburst can be modeled by a $4.5\times 10^7$~yrs-old decaying
starburst 
with a decay time of $5\times 10^6$~yrs and an upper mass cut-off
of $\geq 90\ \mathcal{M}_\odot$, which up to now would have produced
$1.9\times 10^{10}\ \mathcal{M}_\odot$
\citep{1998ApJ...500..147S}.
This nuclear starburst might be associated with
a circum-nuclear molecular ring in $^{12}$CO(1-0), 
\citep{1998ApJ...500..147S, 2004ApJ...609...85S}. 
Furthermore, I~Zw~1 shows indications for an ongoing
minor merger with the companion galaxy marked as 
``western source'' in Fig.~\ref{fig:zwgeneral}. 
This source is found with a velocity shift of only
$(170 \pm 110)\ \mathrm{km\ s^{-1}}$ with respect to the 
systemic velocity of I~Zw~1 and with stellar absorption
features of an old stellar population \citep{2001ApJ...555..719C}.
An ongoing merger with the ``western source'' 
is suggested by an extended 
tidal-tail structure in $\ion{H}{i}$ \citep{1999ApJ...510L...7L} and by
a possible small-scale tidal bridge between the companion and the
north-western spiral arm of I~Zw~1 \citep{2003A&A...405..959S}. 
The bright knot to the north of the disk
(``northern source'' in Fig.~\ref{fig:zwgeneral})
is reported to be a projected foreground star
\citep[e.g.][]{1970ApJ...160..405S, 1982ApJ...257...33S, 1990AJ.....99...37H}.
As a Seyfert galaxy, I~Zw~1 represents 
the prototypical NLS1 
\citep{1985ApJ...297..166O} with narrow
permitted lines, weak $[\ion{O}{iii}]$ emission, 
strong 
$[\ion{Fe}{ii}]$ emission, a
steep soft X-ray spectrum, and with X-ray variability 
\citep{1968ApJ...152L..31S,1976ApJ...208...37P,1979ApJ...230..360O,
1990AJ.....99...37H,
1996A&A...305...53B,2004A&A...417..515V,2004A&A...417...29G}.
The optical spectrum of I~Zw~1 reveals a complex mixture of 
four different line systems, including
a high-excitation narrow-line system 
([\ion{O}{iii}], [\ion{Ne}{iii}], 
[\ion{N}{ii}], [\ion{Fe}{vii}], and [\ion{S}{ii}]) with
$\mathrm{FWHM}\sim 1900\ \mathrm{km\ s^{-1}}$ and blueshifted
by $1450\ \mathrm{km\ s^{-1}}$  \citep{2004A&A...417..515V}. 
Similarly high blueshifts of about $1350\ \mathrm{km\ s^{-1}}$
are exhibited by the NIR coronal 
lines of [\ion{Si}{vi}]~$\lambda =19\,634 $~\AA\ and 
[\ion{Al}{ix}]~$\lambda = 20\,400~$~\AA\ and might be manifestations
of AGN-driven outflows
\citep{1998ApJ...500..147S}. 

An observational case study 
of I~Zw~1 with focus on the implications
of a possible merger-induced star formation and
AGN activity is presented in this paper. 
The data set 
is based on NIR imaging and spectroscopy 
made with ISAAC at the
Very Large Telescope (VLT) of the European Southern 
Observatory (ESO) on Cerro Paranal (Chile) and on
complementary optical imaging made with EFOSC2 at the 
3.6m telescope of
ESO on La Silla (Chile). 
The observations and the reduction methods are described
in Sect.~\ref{sec:obs}.
In Sect.~\ref{sec:results.imaging}, the ISAAC images
of I~Zw~1 are
discussed with a focus on morphology.
Color images of the I~Zw~1 host are 
presented in Sect.~\ref{sec:results.colors}. 
The nuclear spectrum of I~Zw~1 as well as the
spectra of 
the northern and the western source
are analyzed in Sect.~\ref{sec:results.spectra}.
The results
are discussed in Sect.~\ref{sec:discussion} and summarized
in Sect.~\ref{sec:summary}.

\begin{table*}
\caption{Summary of the NIR observations of I~Zw~1 discussed in this
paper.
Column 1: Observation mode,
Column 2: Target (position angle of the long-slit in the case
of the spectroscopic data), 
Column 3: Filter name and central wavelength [$\mu$m],
Column 4: Date of observations, 
Column 5: Detector integration time (DIT), 
Column 6: Number of exposures pre-averaged on the chip (NDIT), 
Column 7: Number of exposures stored (NEXP),
Column 8: Conditions of air mass during the observations.}
\label{tab:observations}
$$
\begin{array}{llrcrrrr}
        \hline
        \noalign{\smallskip}
        \hline
        \noalign{\smallskip}
        \mathrm{Mode} & \mathrm{Target} & \mathrm{Filter}\ \  & 
        \mathrm{Date}
        & \mathrm{DIT\ [s]} & \mathrm{NDIT}& \mathrm{NEXP} 
        & \mathrm{Air\ Mass} \\
        \noalign{\smallskip}
        \hline
        \noalign{\smallskip}
       \mathrm{Broad\ Band} & \mathrm{I\ Zw\ 1} & J,\ 1.250 
        & $2001-08-19$ & 6.000 & 10\ \ \ & 5\ \ \ & 1.263$--$1.268 \\
         \mathrm{Imaging} & \mathrm{GSPC\ S234-E} & J,\ 1.250 
        & $2001-08-19$ & 3.550 & 3\ \ \  & 5\ \ \  & 1.335$--$1.345 \\
             & \mathrm{GSPC\ S234-E} & J,\ 1.250 
        & $2001-08-19$ & 3.550 & 3\ \ \  & 5\ \ \  & 1.394$--$1.405 \\
         & \mathrm{I\ Zw\ 1} & H,\ 1.650 & 
        $2001-08-19$ & 6.000 & 10\ \ \  & 5\ \ \  & 1.258$--$1.260 \\
         & \mathrm{GSPC\ S234-E} & H,\ 1.650 & 
        $2001-08-19$ & 3.550 & 3\ \ \  & 5\ \ \  & 1.346$--$1.356 \\
         & \mathrm{GSPC\ S234-E} & H,\ 1.650 & 
        $2001-08-19$ & 3.550 & 3\ \ \  & 5\ \ \  & 1.406$--$1.418 \\
         & \mathrm{I\ Zw\ 1} & Ks,\ 2.160 & 
        $2001-08-19$ & 3.545 & 20\ \ \  & 5\ \ \  & 1.258$--$1.257 \\
         & \mathrm{GSPC\ S234-E} & Ks,\ 2.160 & 
        $2001-08-19$ & 3.550 & 3\ \ \  & 5\ \ \  & 1.357$--$1.368 \\
         & \mathrm{GSPC\ S234-E} & Ks,\ 2.160 & 
        $2001-08-19$ & 3.550 & 3\ \ \  & 5\ \ \  & 1.419$--$1.432 \\
        \noalign{\smallskip}
        \hline
        \noalign{\smallskip}
       \mathrm{Low\ Resolution} & \mathrm{I\ Zw\ 1\ } (0\degr) & SK,\ 2.200 &
        $2001-07-02$ & ~60.000 & 1\ \ \  & 30\ \ \  & 1.34$--$1.28 \\
             \mathrm{Spectroscopy}   & \mathrm{HD\ 166224} & SK,\ 2.200 & 
        $2001-07-02$ & ~~5.000 & 6\ \ \  & 2\ \ \  & 1.44 \\ 
                                 & \mathrm{HD\ 166224} & SK,\ 2.200 & 
        $2001-07-02$ & ~~5.000 & 6\ \ \  & 2\ \ \  & 1.16 \\
                                 & \mathrm{SAO\ 92129} & SK,\ 2.200 & 
        $2001-07-02$ & ~40.000 & 1\ \ \  & 8\ \ \  & 1.23 \\
                                 & \mathrm{I\ Zw\ 1\ } (43\degr) & SK,\ 2.200 & 
        $2001-09-22$ & ~60.000 & 1\ \ \  & 30\ \ \  & 1.33$--$1.27 \\
                                 & \mathrm{SAO\ 92128} & SK,\ 2.200 & 
        $2001-09-22$ & ~3.550 & 5\ \ \  & 8\ \ \  & 1.40$--$1.38 \\
                                 & \mathrm{SAO\ 92129} & SK,\ 2.200 & 
        $2001-09-22$ & ~10.000 & 4\ \ \  & 8\ \ \  & 1.24$--$1.23 \\
                                 & \mathrm{I\ Zw\ 1\ (43\degr)} & SH,\ 1.650 & 
        $2001-09-22$ & ~70.000 & 1\ \ \  & 24\ \ \  & 1.25$--$1.27 \\
                                 & \mathrm{SAO\ 92128} & SH,\ 1.650 & 
        $2001-09-22$ & ~~3.550 & 5\ \ \  & 8\ \ \  & 1.29$--$1.30 \\
                                 & \mathrm{SAO\ 92129} & SH,\ 1.650 & 
        $2001-09-22$ & ~40.000 & 1\ \ \  & 8\ \ \  & 1.23$--$1.22 \\
        \noalign{\smallskip}
        \hline
        \noalign{\smallskip}
\end{array}
$$
\end{table*}

\section{Observations and Data Reduction\label{sec:obs}}

The NIR program for I~Zw~1 
comprises
imaging in the $J$-, $H$, and $Ks$-bands and
long-slit spectroscopy in the $H$- and $K$-bands.
It
was carried out with ISAAC at the VLT 
from July to September 2001\footnote{The $J$-band images alone 
have been used previously for a 
bulge-disk-decomposition \citep{2003A&A...405..959S}.}. 
The observations relevant for this
paper are listed in Table~\ref{tab:observations}.
The NIR observations are complemented with standard stars used 
for the photometric calibration and/or the telluric correction.
The optical EFOSC2 images, available in the ESO/ST-ECF Science Archive
Facility, are based on observations taken in September 2005.

\subsection{Imaging\label{sec:obs.imaging}}

The reduction of the ISAAC images 
is based on the THELI pipeline
\citep{2005AN....326..432E} and cross-checked 
with independent reductions using IRAF\footnote{Image 
Reduction and Analysis Facility written at the 
National Optical Astronomy Observatories (NOAO) in Tucson, Arizona}
and the DPUSER software for astronomical image
analysis\footnote{Software package written
by Dr. T. Ott, see also \citet{1990ASPC...14..336E}}.
Only the first four frames in each filter contain the 
object at different positions on the array and are used
for the reduction. The fifth frame, which is a pure sky
image, is omitted.
 Before the final 
co-addition of all images, the
THELI reduction includes
(i) flat-fielding, 
(ii) the subtraction of a super-flat,
(iii) a correction for 
the remaining reset anomaly of the array via subtracting
the profile of the image collapsed along the x-axis,
(iv) a bad-pixel identification via weightings of the
flat and the individual science frames, and
(v) sky-subtraction based on sky models 
derived from the background in each individual science 
frame after a 
suitable object masking. 
The Gaussian FWHM of the 
point spread functions
in the reduced images yield 
angular resolutions
of about $0\farcs65$, $0\farcs62$, and $0\farcs46$ for $J$, $H$, and
$Ks$, respectively. 

The night of the imaging observations was classified as photometric
at the beginning. The extinction RMS, however, became more variable
toward the second half of the night when the program for 
I~Zw~1 was executed. 
GSPC~S234-E from the 
catalog of faint
NIR LCO/Palomar NICMOS standards \citep{1998AJ....116.2475P}
was observed twice during the
night. Both observation sequences for this standard were taken 
directly one after 
the other more than one hour before the 
I~Zw~1 observations. This makes it impossible
to crosscheck the stability of the zero-points.
Instead, magnitudes from the 2MASS All-Sky Point
Source Catalog (IPAC Infrared Science Archive (IRSA),
Caltech/JPL)\footnote{http://irsa.ipac.caltech.edu}, which are
available for a few sources in the ISAAC field-of-view,
are used for an alternative cross-check. 
The zero-points measured in the individual sky-subtracted
standard star images one step before co-addition are
$\mathrm{ZP}(J)=25.11$, 
$\mathrm{ZP}(H)=24.75$, and 
$\mathrm{ZP}(Ks)=24.24$. 
These zero-points
are corrected for extinction using 
0.11~mag~(air~mass)$^{-1}$, 
0.06~mag~(air~mass)$^{-1}$, and 0.07~mag~(air~mass)$^{-1}$ in $J$, $H$, and
$Ks$, respectively, as given on the ISAAC web pages.

The optical EFOSC2 images are
reduced via standard reduction 
steps. The calibration in magnitudes is based on
the zero-points logged on the EFOSC2 web pages.
Since the calibration is not completely reliable, the
EFOSC2 images are not used for absolute photometry.

The QSO nucleus is saturated in the ISAAC $H$- and $Ks$-band images
as well as in the EFOSC2 images. The nucleus and the central bulge
region of the QSO host are, therefore, excluded from the analysis of
the imaging data.

\subsection{Spectroscopy\label{sec:obs.spec}}

\begin{figure}
        \centering\resizebox{0.6\hsize}{!}{\includegraphics{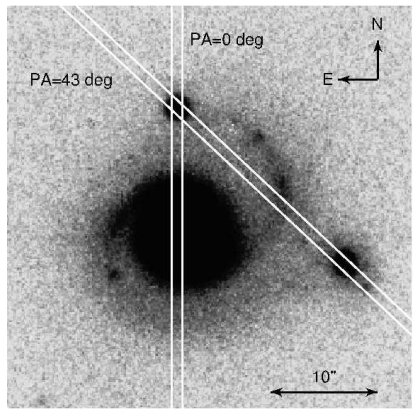}}
        \caption{ISAAC $J$-band image of 
        I~Zw~1 overlaid with the two slit settings used for
          the ISAAC long-slit spectroscopy. The slit for 
	  $PA=0\degr$ 
	includes
          the QSO of I~Zw~1 and the northern source, 
        the slit for $PA=43\degr$ includes the
          northern source and the likely western companion. 
        }
        \label{fig:slitsettings}
\end{figure}

The spectroscopic data on I~Zw~1 were obtained in the 
$H$- and the $K$-bands in 
low-resolution mode ($Rs\approx 600$ for $H$ and
$Rs\approx 500$ for $K$),
using a $1\arcsec$-wide long-slit for settings at position
angles of $PA=0\degr$ and $PA=+43\degr$ 
(Fig.~\ref{fig:slitsettings}).
The setting at $PA=+43\degr$ 
is observed in the $H$- and
$K$-band, the setting at $PA=0\degr$ is 
only observed in $K$.
The $K$-band observations done with the ISAAC 
$SK$ filter cover a wavelength range
of about $7000$~\AA\ ($18\,000$~\AA\ - $25\,000$~\AA) and give 
a dispersion of 7.2~\AA/pix. The FWHM of the arc lines is similar to
$42$~\AA\ which corresponds to $580\ \mathrm{km\ s^{-1}}$. 
The wavelength range and dispersion of the $H$-band observations 
done with the ISAAC $SH$ filter are about $4200$~\AA\ 
($14\,000$~\AA\ - $18\,200$~\AA) and 
4.8~\AA/pix, respectively, which results in a FWHM of the 
arc lines of about $27$~\AA\ or $500\ \mathrm{km\ s^{-1}}$.

The spectroscopic observations are observed with 
a nod-on-the-slit technique. 
The reduction of the 
two-dimensional raw spectra is done using the ISAAC pipeline
recipes and cross-checked with a reduction based on 
IRAF tasks.
The pipeline reduction includes
(i) reducing the spectroscopic flat field by computing
the normalized difference between frames with lamp on and lamp off,
(ii) deriving the slit curvature and wavelength calibration
from arc lamp exposures, (iii) tracing the spectral tilt for different 
positions along the slit by using exposures of a bright
star shifted along the slit, and (iv) the recipe for the final 
reduction of the science and standard star frames. By this recipe,
the frames are flat-fielded
and subsets of the observation sequence
taken at the same position are identified. The frames of each subset
are averaged. These averages are subtracted from each other in both directions
so that each difference also has its negative. After the
correction for slit curvature and spectral tilt and after wavelength
calibration, the subtracted frames are shifted and combined.
From the comparison of the  wavelength solutions of both reductions 
using the argon and xenon arc frames and from comparison with the OH lines,
the maximum error in wavelength is estimated to be 0.1\%.
This accuracy is sufficient for the purpose of this paper.

The contributions of telluric lines and the characteristics of 
the filter curve present in the object spectra are removed by dividing 
by a suitable telluric spectrum. The telluric 
standard stars are reduced exactly
like the object spectra. Since all
standards are G2V or A0V
types, they contain absorption features which are imprinted
as emission artifacts on the object spectrum after telluric correction.
Most prominent are the  
higher-series Brackett features in the $H$-band and the
\ion{Br}{$\gamma$} feature in the $K$-band.
In the case of the nuclear spectrum of
I~Zw~1, three standard star observations are 
available (see Table~\ref{tab:observations}):
HD~166224 - a G2V type star with $K$-band magnitude 7.881
(SIMBAD) - observed at two different air masses during the night, 
and SAO~92129
- a A0V type star with $K$-band magnitude 9.983 (SIMBAD) - observed
once during the night. 
The average spectrum of the two observations of the G2V star
HD~166224
better matches the air mass conditions of the object observation and 
allows a cleaner removal of telluric features (top spectrum in 
Fig.~\ref{fig:zwKspek}). However, compared to the hydrogen
absorptions in the A0V spectrum, the G2V spectrum is characterized by
many more absorption features which 
cause artificial emission lines
in the corrected object spectrum. 
For comparison, both, the spectrum resulting from
the telluric correction and calibration with the average 
spectrum of the two HD~166224 observations and the spectrum resulting
from
the SAO~92129 calibration, are presented.
In the case of the $H$- and $K$-band spectra for $PA=+43\degr$ 
(i.e. the spectra of the northern and western source), the spectra of the
two available A0V standard star observations are averaged before telluric 
correction. Since the telluric 
standard stars frame the object observation in
air mass, this average provides a fair approximation of the air mass
conditions during the object observations. The resulting object spectrum is
compared to a spectrum in which the series of Brackett lines 
is deblended by a Gaussian fit to the 
telluric spectrum before telluric division. In order to enhance the
signal-to-noise ratio (S/N) of the spectrum of the northern source, the spectra
obtained from the $PA=+43\degr$ and the $PA=0\degr$ 
setting are averaged after correction with the A0V 
standard stars.
Artificial emission lines are clearly
marked in the resulting spectra. 
All spectra are extracted in spatial apertures of 3\arcsec\ in diameter.
Absolute flux calibration is done by multiplying the telluric-corrected
object spectra with the black body curves of the 
telluric standard stars,
modeled via the IRAF task STANDARD.

\section{Results\label{sec:results}}

\subsection{Near-Infrared Imaging\label{sec:results.imaging}}

\begin{figure}
        \centering\resizebox{\hsize}{!}{\includegraphics{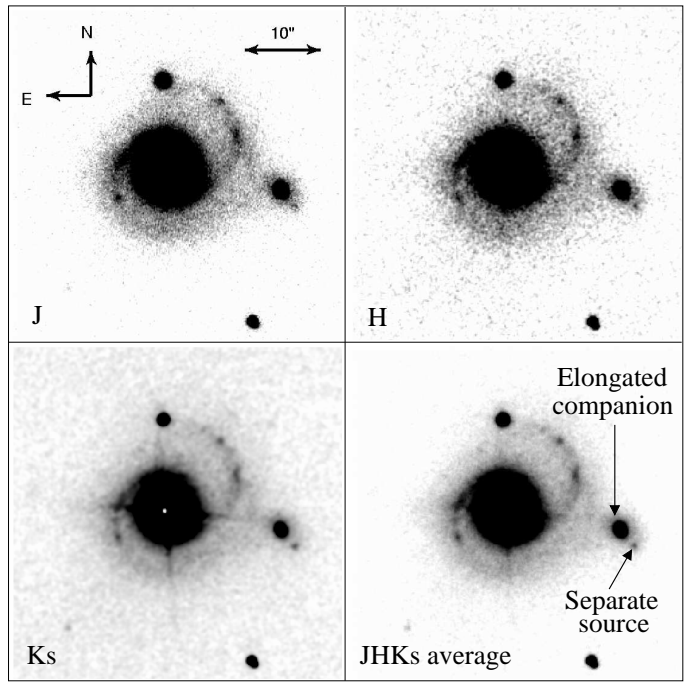}}
        \caption{ISAAC $J$-, $H$-, and $Ks$-band images of 
	I~Zw~1 as well as an average image of all bands. 
	The latter is computed from the images of all three bands
	after they have been smoothed to a similar angular resolution,
	adjusted to a zero background offset, and calibrated in fluxes.
        All
	images are shown in 
        inverse colors and with an arbitrary scaling which pronounces the 
	respective features best. 
	}
        \label{fig:images}
\end{figure}
\begin{figure}
        \centering\resizebox{\hsize}{!}{\includegraphics{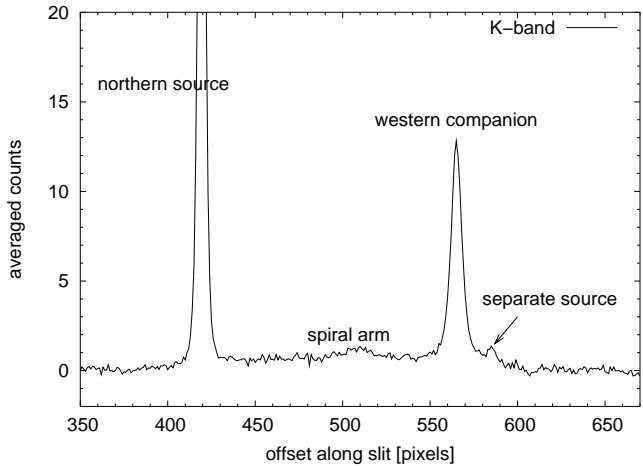}}
        \caption{Cut along the spatial axis of the two-dimensional $K$-band
          spectrum obtained for the $PA=+43\degr$ setting
          (see Fig.~\ref{fig:slitsettings}). The cut consists of an average
          over almost all the spatial lines of the image.
         }
        \label{fig:slitcut}
\end{figure}

 
The images of I~Zw~1 in $J$, $H$, and $Ks$
as well as an image computed from the average of all three bands 
are shown in Fig.~\ref{fig:images}. 
The images 
provide further evidence of a possible tidal bridge toward the 
elongated western companion, as already
indicated by the $J$-band 
image alone \citep{2003A&A...405..959S}.
The image in $Ks$ and the averaged image suggest that 
the feature to the south-west of the companion
resolves into a separate source.
This feature was previously interpreted as a kind
of tidal tail at the far end
of the companion \citep{2003A&A...405..959S}.
The separate source is also 
resolved in the HST image (Fig.~\ref{fig:zwgeneral}). 
The nature of this source
remains unclear. While it might be associated with the
likely companion, it could also be a background or
foreground object. The object displays a red color 
in the $JHKs$ color composite in 
Fig.~\ref{fig:colors} (see Sect.~\ref{sec:results.colors}).
The object is also visible in the spatial cut of the ISAAC spectra 
belonging to the $PA=+43\degr$ slit setting and is
most pronounced in the $K$-band. Fig.~\ref{fig:slitcut} 
shows the average of 
almost all spatial lines of the 
two-dimensional $K$-band spectrum.
However, the S/N of this separate source
is too low for a spectroscopic analysis.

\subsection{Colors\label{sec:results.colors}}

The optical-to-NIR colors of the I~Zw~1
host suggest a concentration of blue stellar populations in the
western part of the host galaxy
and indicate red colors of the sources to the north and west
of the I~Zw~1 disk.

\begin{figure}
        \centering\resizebox{0.4\hsize}{!}{\includegraphics{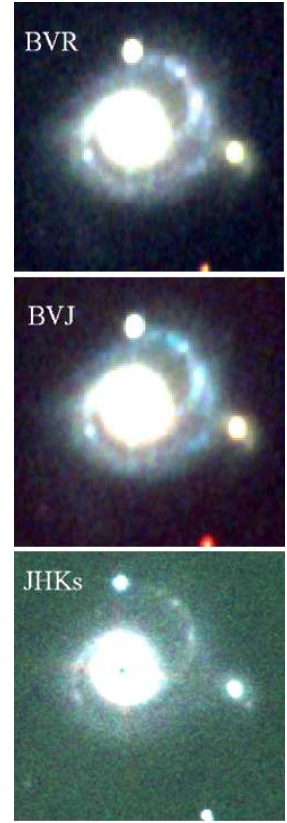}}
        \caption{Optical-to-NIR color composites of I~Zw~1. 
        The order of the filters in the labels represents the order
        of the 
        blue, green, and red channels. 
        The $BVR$ and $BVJ$ images are shown at the EFOSC2 pixel scale of
        0\farcs314 per pixel and an the $B$-band angular resolution of 
	about 1\arcsec.
        The $JHKs$ image is shown at the ISAAC pixel scale of $0\farcs1484$ 
       per pixel and
       an angular resolution of $0\farcs65$.
	See text for more details.
       (The color version of this figure is available in the electronic
         edition.)}
        \label{fig:colors}
\end{figure}

\begin{figure*}
        \centering\resizebox{\hsize}{!}{\includegraphics{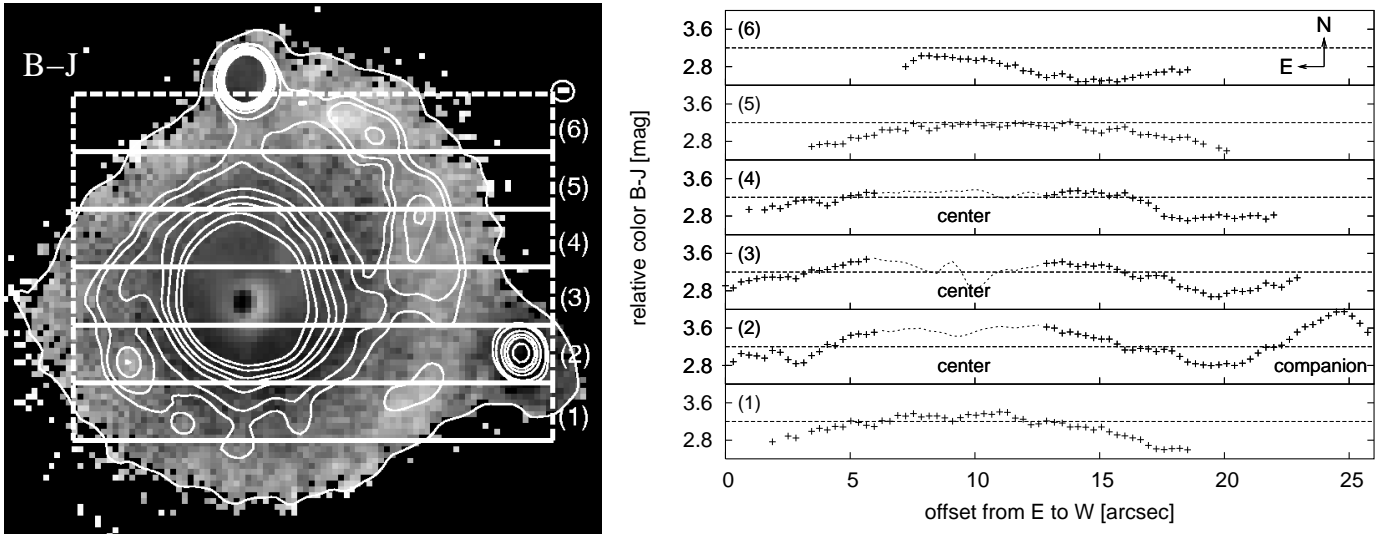}}
        \caption{Relative color trends in $B-J$ for a scan across the host 
         galaxy of I~Zw~1. 
	The left panel shows in grey-scale the $B-J$ image with lighter grey
	indicating bluer colors. 
	The image is shown
	at the EFOSC2 pixel scale of 0\farcs314 per pixel and angular 
	resolution of 1\arcsec.
	It is overlaid by white 
	contours from the 
	$B$-band image.	
	The lowest contour shows the 5$\sigma$ level of the standard deviation
	of the sky in the $B$-band image which is used as a 
	lower limit for pixel rejection (rejected pixels shown in 
	black). 
	For each of the cuts indicated (white boxes) 
	an east-west projection of the mean $B-J$ color is plotted in the
	graph on the right. The color trends are    
	averaged over ten pixels (i.e. about 3\farcs1) 
	in north-south direction (width of the white boxes).
	The central region of I~Zw~1, i.e. approximately 
	the region inside the highest contour level in the left panel,
	is marked by dashed lines
	in projections (2), (3), and (4). This region is affected by
	artefacts from image registration and QSO saturation and not reliable
	for studying color trends.
	See text for more details.}
	\label{fig:colorcuts}
\end{figure*}

The color information is presented in Figs~\ref{fig:colors}
and \ref{fig:colorcuts} in the form of RGB three-color composites
and color projections, respectively.
The three-color images in Fig.~\ref{fig:colors}
are composed of flux-calibrated images using SAOImage
DS9 \citep[see][]{2003ASPC..295..489J}. For each composite,
the color distribution of the 
red, green, and blue input images is uniformly set to
a linear scale with upper and lower limits of 
95\% about the mean of the pixel distribution. Furthermore,
all three images are set to the same bias and contrast values.
The $BVR$ and $BVJ$ images
are shown at the EFOSC2 pixel scale of 0\farcs314 per pixel
and the $B$-band angular resolution of about 1\arcsec.
The $J$-band image is rebinned and smoothed accordingly.
The $JHKs$ image is shown at the ISAAC pixel scale
of $0\farcs1484$ per pixel and the 
$J$-band angular resolution of $0\farcs65$.
A grey-scale representation of the $B-J$ image together
with a rough scan of $B-J$ projections across the I~Zw~1
host galaxy is 
presented in Fig.~\ref{fig:colorcuts}.
In this Figure, all pixels with
values lower than 5$\sigma$ of the sky standard deviation in the
$B$-band image are neglected. As the scans are intended to mainly highlight
east-west color trends, each measurement is 
averaged over ten pixels in north-south
direction. By this method, north-south structures are
traced well, while north-south gradients close to the central region are 
diluted. 

The concentration of blue color in the western part
of the I~Zw~1 host galaxy
is most obvious in the $BVJ$ composite but is
also noticeable in $BVR$
(Fig.~\ref{fig:colors}). In particular, the putative tidal
bridge between the north-western spiral arm and the
companion galaxy (see Sect.~\ref{sec:results.imaging})
is characterized by blue color.
The blue color asymmetry between the eastern and the western
part of the I~Zw~1 host galaxy
has not explicitly been noted before. Retrospectively, the
effect seems weakly visible
in the $B$- and $I$-band
composite presented by \citet[][ Fig.~2 therein]{2001AJ....122.2791S}.
The much weaker evidence for the
color asymmetry in this $BI$ composite compared to
Fig.~\ref{fig:colors} is probably due to the fact
that the $BI$ composite is based on a linear interpolation 
from the $B$- and $I$-band data and
plotted with less contrast on the spiral arm region. 
A quantitative analysis of the blue color asymmetry in the
I~Zw~1 host is presented
in Fig.~\ref{fig:colorcuts}.
The projections in the right panel of this Figure show that there
is only one prominent blue region in the eastern part of the
host galaxy, i.e. the blue knot seen in projection (2).
On the contrary, the western part of the host galaxy 
is characterized by similarly blue regions in all of the
six projections.
The blue color suggests an enhanced level
of star formation activity. Enhanced star formation activity
in this particular part of the I~Zw~1 host
close to the putative companion
galaxy might be causally connected to the
supposed minor merger process. It is likely that the star formation
is directly induced by the tidal interaction with the
western companion galaxy.
Enhanced star formation activity in the 
north-western spiral arm has been 
discussed in previous publications. 
\citet{2001ApJ...555..719C} conclude 
that the emission lines
in the spectrum of the north-western spiral arm are
typical of \ion{H}{ii} regions and indicative of 
ongoing star formation. Similarly, 
\citet{1998ApJ...500..147S} find star formation
activity in the north-western spiral arm
by population synthesis for a region about
9\arcsec\ west and 5\arcsec\ north of the nucleus of
I~Zw~1. They also report this region to be 
associated with a significant amount of molecular
gas.

In the NIR $JHKs$ composite, the 
I~Zw~1 host shows a uniform color.
As a NIR composite, this color image mainly traces the 
old stellar population underlying throughout the 
host galaxy.
The analysis of the colors of the QSO nucleus and 
the central region of the I~Zw~1 host 
is beyond the scope of the data discussed
here, as
saturation
of the QSO nucleus affects the EFOSC2 images and
the ISAAC $H$- and $Ks$-band images. 
Nevertheless, the outskirts of the bulge show
indications for a red color in the
color images (Fig.~\ref{fig:colors}) and
in the $B-J$ projections across the nucleus
(Fig.~\ref{fig:colorcuts}). 
Such a red color
might be an indication of an older 
stellar population. However, the putative circum-nuclear
starburst \citep{1998ApJ...500..147S}
and the low mean $J$-band mass-to-light
ratio of the bulge \citep{2003A&A...405..959S}
rather 
suggest a young stellar population. 
Alternatively,
the reddening could be caused by an increased amount
of dust extinction. Dust is suggested by the 
HST image in Fig.~\ref{fig:zwgeneral} which shows
some dust patches in the stellar bulge.
Correspondingly, previous NIR studies of I~Zw~1
suggest a visual extinction of about 10~mag for the 
stellar component in the
central region of the host galaxy
\citep{1994ApJ...424..627E, 1998ApJ...500..147S}.
Finally, as a consequence of the low angular resolution,
the red colors might still be dominated by emission
from the nucleus. The nuclear colors of I~Zw~1
as seen in the bands $B$, $I$, $H$, and $K'$ have 
previously been reported to show a NIR excess
\citep{1999ApJ...512..162S}.
According to Figs 6 and 7 in \citet{1999ApJ...512..162S}, this
excess
can mostly be explained by a hot dust component combined
with a QSO spectral energy distribution.
Higher angular resolution via adaptive optics or
interferometric observations will be needed 
for a more precise study of 
color gradients close to the 
center of I~Zw~1.

The northern source and the western source are 
relatively red
objects in each of the color composites in
Fig.~\ref{fig:colors} (see also 
Fig.~\ref{fig:colorcuts}, in projection (2) for the western source). 
The red color
of the likely northern foreground star agrees with the 
late stellar type discussed in Sect.~\ref{sec:results.spectra}. 
The red color of the
western companion supports the scenario
of a predominantly
old and evolved stellar population (see 
Sect.~\ref{sec:results.spectra}). The separate object
to the south-west of the likely western companion 
(see Sect.~\ref{sec:results.imaging}) appears
as a red object in the $JHKs$ composite.

\subsection{Near-Infrared Spectra\label{sec:results.spectra}}

\begin{table*}
        \caption{Measured fluxes of the lines 
identified in the nuclear $K$-band spectrum 
of I~Zw~1 (upper spectrum in Fig.~\ref{fig:zwKspek}). Values in brackets are uncertain measurements only listed for completeness. Column~1: Line and rest wavelength, Column~2: Comment about the component of the line, Column~3: Gaussian FWHM, Column~4: Observed wavelength, Column~5: Flux measured from the ISAAC spectrum, Column~6: Flux
published by \citet{1998ApJ...500..147S}.}
        \label{tab:Klines}
        $$
        \begin{array}{llcccc}
        \hline
        \noalign{\smallskip}
        \hline
        \noalign{\smallskip}
\mathrm{Line} & \mathrm{Component} 
   & \mathrm{FWHM}^{\mathrm{a}} & \mathrm{\lambda_{obs}} 
   & \mathrm{Flux}^{\mathrm{b}} 
   & \mathrm{Flux\ (Schinnerer\ et\ al.\ 1998)}\\
              &              
   & \mathrm{[km\ s^{-1}]} & [$\AA $] 
   & \mathrm{[10^{-15}\ erg\ s^{-1}\ cm^{-2}]} 
   & \mathrm{[10^{-15}\ erg\ s^{-1}\ cm^{-2}]} \\ 
        \noalign{\smallskip}
        \hline
        \noalign{\smallskip}
\multicolumn{5}{l}{\mathbf{Permitted\ lines\ with\ narrow\ and\ broad\ components}} \\
$\ion{Pa}{$\alpha$}$\ \lambda 18\,756   
   & \mathrm{narrow} & 920 &  & 86   &  \\
                                
   & \mathrm{broad} & 3110\enspace &            & 107\enspace     &  \\
                                         
   & \mathrm{total} &           &    19\,895        & 193\enspace    & 284\pm 12\enspace\\
$\ion{Br}{$\delta$}$\ \lambda 19\,451  
   & \mathit{(narrow} & \mathit{920} &   & \mathit{11)}\enspace  &    \\
 
   & \mathit{(broad} & \mathit{3110}\enspace &              & \mathit{7)}  &  \\
                                       
   & \mathrm{total} &           &     20\,639    & 18  & 3.7\pm 1.8\\
$\ion{Br}{$\gamma$}$\ \lambda 21\,661   
   & \mathit{(narrow} & \mathit{920} &  & \mathit{11)}\enspace  & \\

   & \mathit{(broad} & \mathit{3110}\enspace &           & \mathit{6)}  &            \\
                                         
   & \mathrm{total} &           &  22\,971   & 17 & 26\pm 8\enspace \\
        \noalign{\smallskip}
        \hline
        \noalign{\smallskip}
\multicolumn{5}{l}{\mathbf{Narrow\ forbidden\ lines\ with\ blueshift}} \\
$[\ion{Si}{vi}]$\ \lambda 19\,634 & \mathrm{blue}     & 920
         & 20\,732 &  7 &  3.5\pm 1.8\\ 
\mathit{(+ H_2\ 
%\nu=1-0\ 
S(3)\ \lambda 19\,576}   & \mathit{systemic} &  
         & \mathit{\sim 20\,769)}\enspace\enspace &     &  \\
$[\ion{Al}{ix}]$\ \lambda 20\,400  
%   & \mathrm{systemic} & 1228.9\pm & 21672.2 & 5.43\pm  &  \\

   & \mathrm{blue}     & & \mathrm{not\ detected}  &   &  3.2\pm 3.2\\
\mathit{(+ H_2\ 
%\nu=1-0\ 
S(2)\ \lambda 20\,338}   & \mathit{systemic} &  
         & \mathit{\sim 21\,578)}\enspace\enspace &     &  \\
        \noalign{\smallskip}
        \hline
        \end{array}
        $$
\begin{list}{}{}
\it{\item[$^{\mathrm{a}}$] The spectral resolution is $\mathit{650\ km\ s^{-1}}$.}
\it{\item[$^{\mathrm{b}}$] The $1~\sigma $-error of the flux is $\mathit{1 \times 10^{-15}\ erg\ s^{-1}\ cm^{-2}}$.}
\end{list}
\end{table*}
\begin{figure*}
        \centering\resizebox{0.9\hsize}{!}{\includegraphics{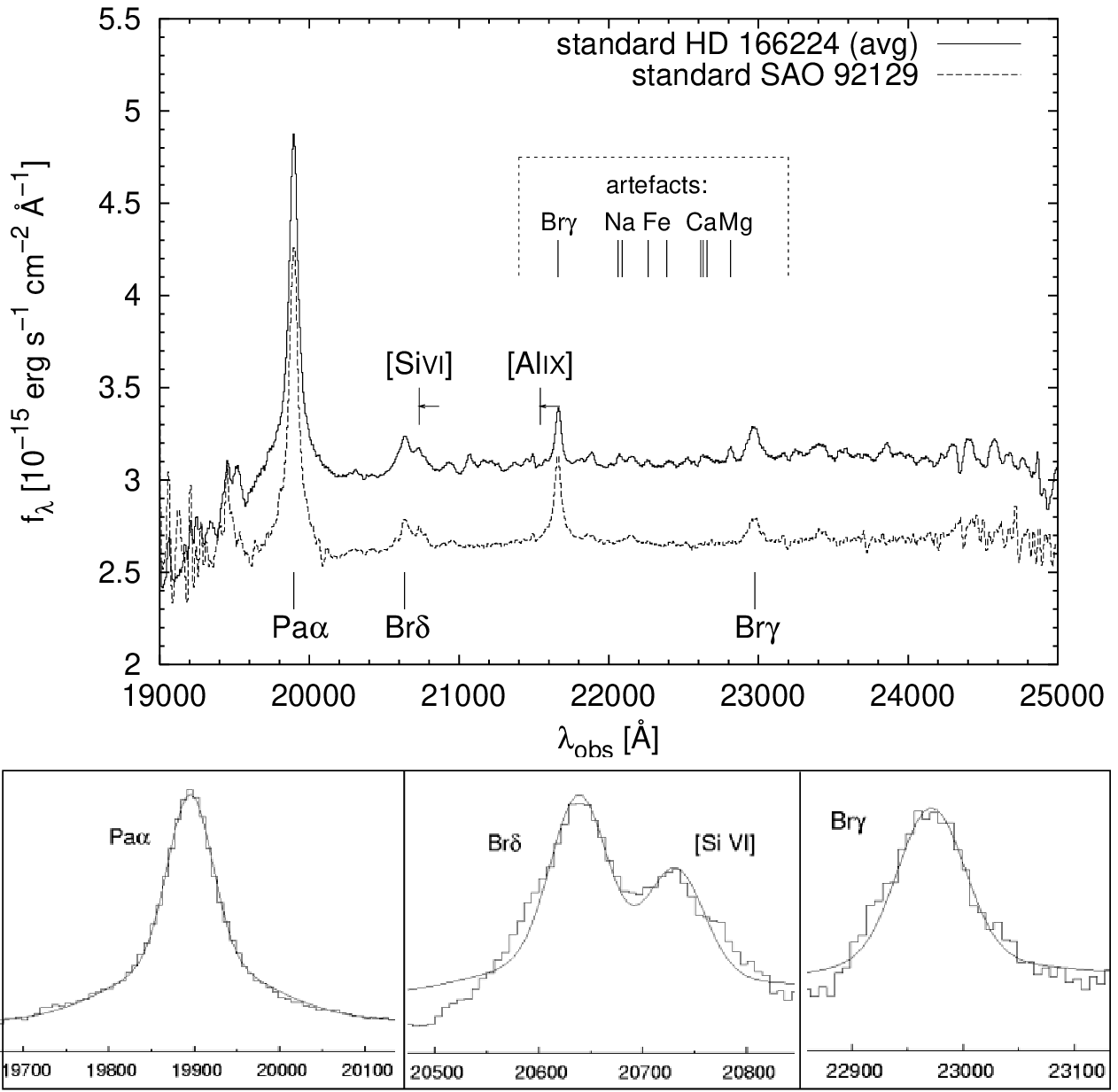}}
        \caption{ISAAC $K$-band spectrum of the nuclear region of 
        I~Zw~1 (upper panel) and results for the 
        individual lines from fitting the 
        HD~166224-corrected spectrum (lower panels). 
        For a better identification of telluric correction
	artifacts, the full spectrum is shown as corrected and calibrated
        by (i) using the average of the two observations of the
	G2V star HD~166224 and (ii) using the observation
	of the A0V star SAO~92129. 
	The spectrum is extracted in a $3\arcsec$-wide aperture 
        centered on the 
	QSO nucleus and boxcar-smoothed to a resolution of 
        about 
	$47$~\AA\ ($\sim \mathrm{650\ km\ s^{-1}}$).
	Prominent lines and telluric-correction artifacts 
        are marked at the respective observed wavelengths.
	$[\ion{Si}{vi}]$ and $[\ion{Al}{ix}]$ as well as the arrows 
        visualizing the blueshift are marked according to the positions
        reported by \citet{1998ApJ...500..147S}.}
        \label{fig:zwKspek}
\end{figure*}
The nuclear spectrum of I~Zw~1 is 
plotted in Fig.~\ref{fig:zwKspek}. 
Since the night was not completely photometric, the absolute flux calibration
implies uncertainties. The independent calibrations with the 
two standard stars observed during the same night
yield a difference in flux density of 
about 20\% (Fig.~\ref{fig:zwKspek}). However, 
within this uncertainty, the mean flux density 
of the flux-calibrated nuclear spectrum is
$2.9\times 10^{-15}\ \mathrm{erg\ s^{-1}\ cm^{-2}}$~\AA$^{-1}$ or
$0.049\ \mathrm{Jy}$ and corresponds 
to a 2MASS $Ks$ magnitude\footnote{A $K$-band flux zero-point 
of 666.8~Jy is adopted \citep{2003AJ....126.1090C}.} 
of $10.4$~mag. This
agrees with the cataloged 2MASS $Ks$ magnitude of I~Zw~1 of
$10.35$~mag (2MASS All-Sky Point-Source Catalog).
Previous $K$-band spectra of I~Zw~1,
measured within
similar extraction apertures, are reported to be flux-calibrated by using
aperture magnitudes from the literature
\citep{1998ApJ...500..147S,2001ApJS..136...61S}. The absolute flux of these
spectra is larger than the ISAAC flux 
by about a factor of 1.5. This corresponds to a magnitude
difference of about 0.4. A similar magnitude difference is found in the
ISAAC imaging data. The nuclear $J$-band magnitude of I~Zw~1,
as derived from the ISAAC image, agrees with the 2MASS measurement but is
by about 0.3 to 0.4~mag fainter than a $J$-band measurement taken 
about nine years earlier \citep{1994ApJ...424..627E}.
While the ISAAC data are saturated in $H$ and $Ks$, the corresponding
measurements from 2MASS show a similar offset with respect to the 
earlier data by \citet{1994ApJ...424..627E}.
If not a calibration offset, this effect might be a manifestation of the
NIR variability of I~Zw~1 
\citep[see e.g.][]{1989AJ.....97..957N, 1999AJ....118...35N}.

The results from fitting the continuum and the lines in the nuclear
ISAAC spectrum of I~Zw~1 are shown in Table~\ref{tab:Klines}. 
The fit is done for the upper spectrum of Fig.~\ref{fig:zwKspek} 
by means of the software 
Specview\footnote{Specview is a product of the Space Telescope Science 
Institute, which is operated by AURA for NASA.\label{foot:specview}}.
The continuum is modeled using a power-law component 
$f_\lambda = f_{\lambda _\mathrm{ref}}(\lambda/\lambda_{\mathrm{ref}})^\alpha$
with $\alpha = 0.183$, $\lambda_{\mathrm{ref}}=21\,964$~\AA, and
$f_{\lambda _\mathrm{ref}}=3.094\times 10^{-15}\ 
\mathrm{erg\ s^{-1}\ cm^{-2}}$~\AA$^{-1}$.
The hydrogen lines are modeled via
the superposition of a broad and a narrow Gaussian component (lower
panels in Fig.~\ref{fig:zwKspek}). This two-component model is suggested by
the shape of the highest S/N hydrogen line $\ion{Pa}{\alpha}$. Only the
$\ion{Pa}{\alpha}$ line 
is used to determine the FWHMs of both Gaussian components.  
For the fits of the low S/N lines $\ion{Br}{\delta}$ and 
$\ion{Br}{\gamma}$, the FWHMs are 
taken as fixed parameters. 
Line centers are determined by eye, and both Gaussian components of each
line are constrained to the same center and fitted simultaneously.
Out of the two blueshifted high-excitation lines reported by 
\citet{1998ApJ...500..147S}, only $[\ion{Si}{vi}]$ is detected on the 
red wing of the $\ion{Br}{\delta}$ line and modeled with the narrow
Gaussian component. 
$[\ion{Si}{vi}]$ is additionally blended
with $H_2\ \nu=1-0\ S(3)$ at the systemic velocity of I~Zw~1.
This contributes to the uncertainty of the flux measurement for
$[\ion{Si}{vi}]$ reported in Table~\ref{tab:Klines}. As
$H_2\ \nu=1-0\ S(1)$ at $\lambda = 21\, 218$~\AA\ seems to be below the
detection limit in Fig.~\ref{fig:zwKspek} and assuming that
the line strength of $H_2\ \nu=1-0\ S(3)$ is less or similar 
\citep{1990ApJ...352..433K}, the contribution of $H_2\ \nu=1-0\ S(3)$
to the $[\ion{Si}{vi}]$ flux is likely to be small.
 
The split into two Gaussian components is less
certain for the Brackett lines. The corresponding 
values are parenthesized in
Table~\ref{tab:Klines}. 
The fluxes determined for the $\ion{Pa}{\alpha}$ and 
$\ion{Br}{\gamma}$ lines are by a factor of about 1.5
lower than the fluxes reported by \citet{1998ApJ...500..147S}.
This is the same factor as found for the disagreement in 
continuum flux densities. The measured flux of
the $\ion{Br}{\delta}$ line shows a different behavior in
comparison to the data by \citet{1998ApJ...500..147S}. This 
measurement, however, is the most uncertain one because
of the blending with $[\ion{Si}{vi}]$ and a possible
small contribution from 
$H_2\ S(3)$. Line ratios (see below)
seem to indicate that the new measurement based on the 
ISAAC data is closer to reality.
The RMS of the background flux density outside the obvious line emission from
I~Zw~1 and outside the $\ion{Br}{\gamma}$ artifact 
is estimated as 
$3\times 10^{-17}\ \mathrm{erg\ s^{-1}\ cm^{-2}}$~\AA$^{-1}$. 
This is a conservative compromise between the less noisy background in the 
central part of the band and the increasing noise toward the band limits.
As an effect of the stronger correction artifacts in the one case and the 
stronger telluric residuals in the other case, the G2V- and the A0V-corrected
spectrum show a similar background RMS.
Multiplied by the
spectral resolution element the given flux density RMS corresponds to a 
$1\ \sigma$ flux error of about 
$1\times 10^{-15}\ \mathrm{erg\ s^{-1}\ cm^{-2}}$.

The line centers measured for the three hydrogen lines result in
a redshift of $z\approx 0.06$. This agrees with the redshift reported for
I~Zw~1 in the literature \citep[e.g.][]{1985AJ.....90.1642C}.
The two FWHMs of the narrow and the broad Gaussian component are similar
to the fit of the hydrogen lines in the optical spectrum of I~Zw~1
reported by \citet{2004A&A...417..515V}. These authors use a narrow Lorentzian 
component with an 
FWHM of $1100\ \mathrm{km\ s^{-1}}$ and report the need for an additional
broad Gaussian component with an FWHM of $5600\ \mathrm{km\ s^{-1}}$.
The putative line center of the $[\ion{Si}{vi}]$ feature displays 
a blueshift of
about $1460\ \mathrm{km\ s^{-1}}$ with respect to the velocity 
of the hydrogen line system of I~Zw~1. 
\citet{1998ApJ...500..147S} report a blueshift of 
$1350\ \mathrm{km\ s^{-1}}$ for the two high-excitation lines
$[\ion{Si}{vi}]$ and $[\ion{Al}{ix}]$.
\citet{2004A&A...417..515V} speculate that
the NIR high-excitation lines may originate from the 
two high-excitation narrow-line
systems, N1 and N2, which they
found by fitting the optical spectrum of I~Zw~1. 
The lines
belonging to these systems at optical wavelengths, i.e.
[\ion{O}{iii}], 
[\ion{Ne}{iii}], and [\ion{N}{ii}]
(plus [\ion{Fe}{vii}] and [\ion{S}{ii}] in the case of N1), 
are relatively broad and blueshifted by up to 
$1450\ \mathrm{km\ s^{-1}}$ in N1. The existence of 
a blueshifted system of [\ion{O}{iii}] and 
[\ion{Ne}{iii}] lines has also been reported
in previous publications \citep{1976ApJ...208...37P,1979ApJ...230..360O,
1997ApJ...489..656L}. Similar to other QSOs, the spectrum of I~Zw~1 
shows a trend of increasing blueshift for lines of increasing ionization
level \citep{1997ApJ...489..656L}.

The line ratios computed for the hydrogen lines of 
Table~\ref{tab:Klines} indicate conditions similar to those in typical 
$\ion{H}{ii}$ regions, possibly affected by a certain amount of 
dust reddening.
This is suggested by the fact that
the ratios $\ion{Pa}{\alpha}$/$\ion{Br}{\gamma}\sim 11$ and 
$\ion{Pa}{\alpha}$/$\ion{Br}{\delta}\sim 11$ are marginally lower than
the theoretical values. Assuming case B recombination 
for a hydrogen plasma
at a temperature of $10^4$~K and a density of
$10^4\ \mathrm{cm^{-3}}$ \citep{1987MNRAS.224..801H}, the 
theoretical values result in 
$\ion{Pa}{\alpha}$/$\ion{Br}{\gamma}=12.7$ and 
$\ion{Pa}{\alpha}$/$\ion{Br}{\delta}=18.34$.
These above line ratios are only computed for the total flux of
each line because of the uncertainties implied in the two-component 
model for the Brackett lines. 
For the ratio of $\ion{Pa}{\alpha}$ to $\ion{Br}{\gamma}$ in the 
nuclear spectrum of I~Zw~1,
\citet{1998ApJ...500..147S} report a similar value of 11:1.
Based on their flux measurements, however, the ratio of 
$\ion{Pa}{\alpha}$ to $\ion{Br}{\delta}$ results in $\sim 77$. 
This is unusually
high compared to the theoretically expected values. Accordingly,
the higher flux of $\ion{Br}{\delta}$ measured in the ISAAC spectrum
might be more realistic.

%=============> Companion
\begin{figure*}
        \centering\resizebox{\hsize}{!}{\includegraphics{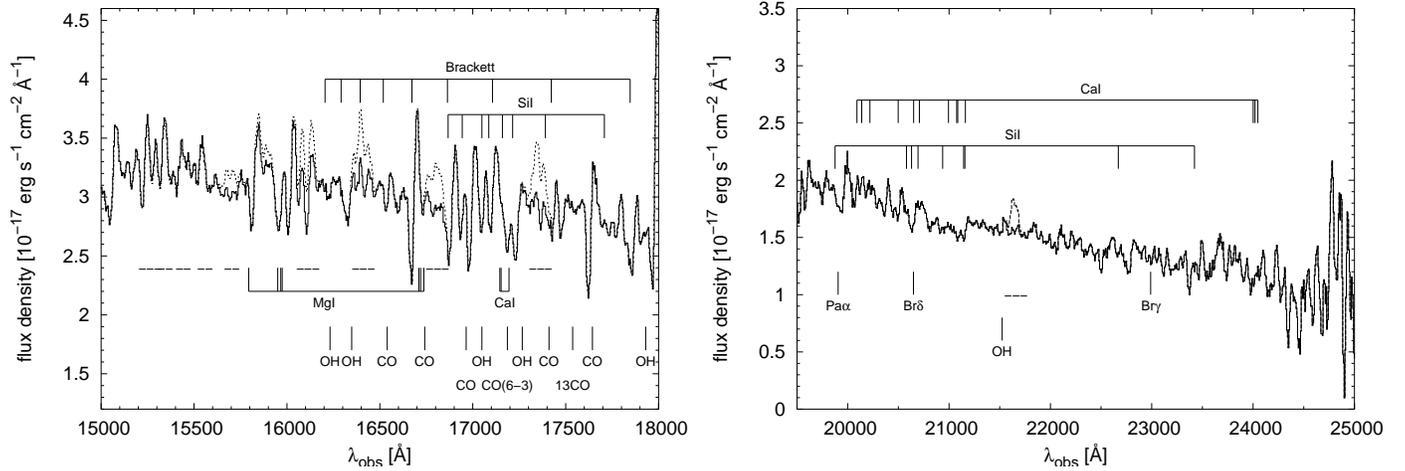}}
        \caption{$H$-band (left panel) and $K$-band (right panel) spectrum of the
        likely western companion galaxy, both extracted in $3\arcsec$-wide
       apertures. The $H$-band spectrum is shown at a resolution of
       $36$~\AA\ ($\sim 670\ \mathrm{km\ s^{-1}}$), the $K$-band spectrum 
       at a resolution of 55~\AA\ 
          ($\sim 770\ \mathrm{km\ s^{-1}}$). The solid curve shows the spectrum after
       correction with an A0V telluric spectrum from which the Brackett 
      lines have
      been deblended. 
      The overlaying thin curve shows the same spectrum corrected
      without previous deblending. Typical lines in this wavelength
      region are marked in the plot at a redshift of  
       $z=0.0616$, i.e. the redshift measured for the
     likely companion by \citet{2001ApJ...555..719C}. 
      The horizontal dashed markers
      indicate the wavelength regions contaminated with artificial 
      hydrogen emission
      in consequence of the telluric correction with A0V standards. 
         }
        \label{fig:compspek}
\end{figure*}
Although less conclusive in consequence of the low S/N, 
the $H$- and $K$-band spectra of the western companion 
(Fig.~\ref{fig:compspek}) seem to
support the findings by \citet{2001ApJ...555..719C} that the companion shows
the spectral properties of an old stellar population.
The lines in Fig.~\ref{fig:compspek} are marked at the
redshift of $z=0.0616$ derived for the likely 
companion by \citet{2001ApJ...555..719C}. Many of these lines
match with
absorption features in the spectra. In particular, the hydrogen
lines are found in absorption rather than emission. This
suggests a lack of star formation activity in the western companion.

%==========> northern source
\begin{figure*}
        \centering\resizebox{\hsize}{!}{\includegraphics{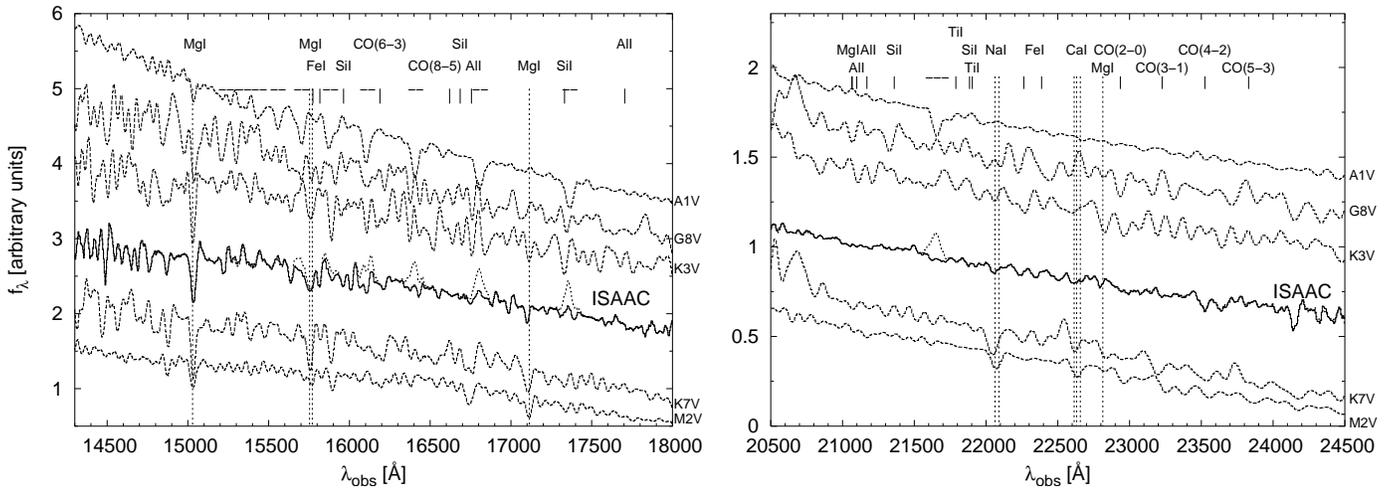}}
        \caption{$H$-band (left panel) and $K$-band (right panel) spectrum of 
        the
        northern source, both extracted in $3\arcsec$-wide
       apertures and compared to stellar templates (see labels)
	from the library by \citet{1992A&AS...96..593L}. 
	The ISAAC spectrum (solid curve) 
	is derived from the telluric correction with an A0V spectrum with 
	deplended Brackett lines. It is overlaid with the spectrum 
	corrected without deblending (thin curve). The spectral resolutions
	are $36$~\AA\ ($\sim 670\ \mathrm{km\ s^{-1}}$) and 55~\AA\ 
	($\sim 770\ \mathrm{km\ s^{-1}}$) in $H$ and $K$, respectively. 
	The library spectra have similar spectral resolutions. For the 
	purpose of comparison with the ISAAC spectra, the 
	available units of wavenumber (in $\mathrm{cm}^{-1}$) versus 
        arbitrary units of flux per unit frequency are
	converted into wavelength 
        (in \AA) versus flux per unit wavelength. 
	All spectra are normalized to
	a flux of 1 at $21\,500$~\AA\ and shifted along the flux-axis for 
	this representation. 
	Typical lines in this wavelength region are labeled. The lines 
	identified in the 
	ISAAC spectra are marked by a 
	vertical dashed line. The horizontal dashed lines indicate the 
	Balmer lines which artificially
	appear as emission lines in the ISAAC spectra as a consequence 
	of the telluric correction 
	via A0V stars. 
         }
        \label{fig:starspeks}
\end{figure*}
The $H$- and $K$-band spectra obtained for the northern source
suggest that the source is a cool star at zero
redshift and most likely a K-type main-sequence 
star. This reconfirms the scenario of a foreground star
\citep{1970ApJ...160..405S, 1982ApJ...257...33S, 1990AJ.....99...37H}, 
contrary to the alternative scenario of a 
progenitor nucleus of a galaxy
interacting with I~Zw~1 or a tidal dwarf galaxy
\citep{mp_aoworkshop_davies05}.
The late type of the star is suggested by 
Fig.~\ref{fig:starspeks} which demonstrates that the
$H$- and $K$-band spectra of the star
agree with the template spectra of local cool stars 
taken from the library by \citet{1992A&AS...96..593L}.
Besides the prominent agreement in $\ion{Mg}{i}$ absorptions, the 
spectrum of the northern source reflects the continuum shape of
cool stars and, in particular, the 
depression shortward of $16\,000$~\AA. The latter is 
attributed to a combined effect of an opacity minimum of 
$\ion{H}{^-}$ and water absorption in the atmospheres of cool stars
\citep{1992A&AS...96..593L}.

\section{Discussion\label{sec:discussion}}

\subsection{Merger History of I~Zw~1}
The consensus from numerical simulations is that 
major mergers typically result in elliptical remnants
\citep[e.g.][]{1992ApJ...400..460H, 2003ApJ...597..893N, 
2005MNRAS.357..753G, 2006MNRAS.369..625N}. The
clear spiral structure of I~Zw~1 is, therefore, an argument
against a recent major merger event. Instead, the evidence for
a minor merger between I~Zw~1 and the 
low-mass western
companion galaxy is increasing
\citep{1999ApJ...510L...7L,2001ApJ...555..719C, 2003A&A...405..959S}.
The data of the present paper indicate a tidal 
interaction via the  
concentration of blue color in the western part of the
I~Zw~1 host galaxy, i.e. the part which is adjacent to the
western source. This leads to the speculation that an ongoing
tidal interaction might manifest itself by enhanced star formation 
activity in those regions of the I~Zw~1 host which
are facing the western source. Merger-enhanced star formation 
in the spiral host of I~Zw~1 is not unexpected, regarding
that the host is suggested to be rich in gas. Based on 
CO observations, \citet{1998ApJ...500..147S} estimate a total
molecular gas mass of about $8\times 10^9\ \mathcal{M}_\odot$ to 
$10^{10}\ \mathcal{M}_\odot$. Furthermore, enhanced star formation activity
has previously been found in the nuclear region and in the
north-western spiral arm 
\citep{1998ApJ...500..147S, 2001ApJ...555..719C}. 

\subsection{Nature of the Western Source}
As discussed in Sections~\ref{sec:results.colors} and 
\ref{sec:results.spectra}, 
the western source does not show any signs for recent 
star formation. 
This lack of recent
star formation in the western source 
does not contradict the scenario that the source is in
an ongoing merger with I~Zw~1.
It rather suggests that the companion galaxy does not have
a sufficient gas reservoir to incite a starburst, as would
be typical for a 
gas-poor dwarf elliptical.
A less plausible alternative is the scenario of a tidal dwarf
galaxy. Tidal dwarf galaxies are young dwarf galaxies formed out of 
the tidal debris
of a recent major interaction. They
usually have a large gas content and young stellar populations
\citep[e.g.][]{2004A&A...427..803D}. This is in contrast to what is 
observed for the companion of I~Zw~1. Furthermore, 
tidal dwarfs are typically born along the extended tidal
tails formed during a major interaction
\citep[e.g.][]{2001A&A...378...51B}. As discussed before, a recent
major interaction between I~Zw~1 and a similar-sized galaxy
is quite unlikely considering the regular spiral structure of 
the I~Zw~1 host.
The above reasoning cannot rule out the possibility that the western companion
is an aged
tidal dwarf galaxy, i.e. a dwarf galaxy of tidal origin. 
A tidal origin of dwarf spheroidals
around the Milky Way or Andromeda has recently been suggested
regarding that
dwarf galaxies may appear as dwarf spheroidals after a Hubble time
of evolution in the dark matter halo of the primary galaxy
\citep{2007astro.ph..1289M}.

\subsection{Nuclear Activity of I~Zw~1}

As an infrared-excess Palomar Green QSO, 
I~Zw~1 has been discussed  
as a possible
QSO at a transition stage in the putative
evolutionary sequence from ULIRGs to QSOs
\citep[e.g.][]{1988ApJ...325...74S, 2001AJ....122.2791S,
2001ApJ...555..719C}. According to this scenario
the QSO activity in I~Zw~1 would be
young. A young stage of the AGN in I~Zw~1
is also suggested via a different line of 
argument: 
Similarities between NLS1s and broad-absorption line QSOs
gave rise to the 
hypothesis that at least those NLS1s with outflows could be 
counterparts of high-redshift, high-luminosity broad-absorption line QSOs 
\citep[e.g.][]{1997ApJ...489L..25L,2000NewAR..44..461B, 2000MNRAS.314L..17M}. 
As one alternative, both classes of objects could be observed
during a young phase of AGN evolution, characterized by smaller
black hole masses and higher accretion rates
\citep[e.g.][]{1995MNRAS.277L...5P, 2000MNRAS.314L..17M,
2003ApJ...593L..11Y, 2004A&A...420..889M, 2004AJ....127.3168B}.
In fact, I~Zw~1 is a NLS1 whose similarity with 
broad-absorption line QSOs has previously been noted.
The strong optical \ion{Fe}{ii} emission of I~Zw~1, 
the weak [\ion{O}{iii}] emission, the "red" UV continuum,
and the strong infrared emission remind of 
low-ionization broad-absorption line QSOs 
\citep{1997ApJ...489..656L}. Furthermore, the blueshifts
of high-excitation lines in the optical and NIR spectra,
as reported by \citet{1998ApJ...500..147S}, 
\citet{2004A&A...417..515V}, and in the present paper,
suggest an AGN-driven outflow.
Although these arguments suggest a young stage
of the AGN in I~Zw~1, it cannot be excluded that the
AGN might just appear young.

Furthermore, it remains unclear which process
is responsible for the nuclear activity
in I~Zw~1 and which part is assigned to the
tidal interaction with 
the western companion galaxy.
The conclusions from minor-merger simulations 
by \citet{1995ApJ...448...41H}
suggest that the companion 
galaxy of I~Zw~1 might still be too distant to be
the cause of the AGN in I~Zw~1.
The simulations rather indicate that a minor merger may induce 
central activity in the primary galaxy during the late stages of
the merger, i.e. when the satellite galaxy has reached the central
few kiloparsecs of the primary galaxy \citep{1995ApJ...448...41H}.
Disk star formation, on the contrary, may well be induced
during the early stages of the merger. Thus, the disk star formation 
indicated by the blue colors in the western part 
of the I~Zw~1 host galaxy might indeed be directly connected
to the interaction with the western companion.
The role of tidal interactions in the formation of AGN
in Seyfert galaxies is generally controversial
\citep[see review by][]{2004IAUS..222..395S}.
\citet{2000ApJ...536L..73C} suggests that
instead of triggering nuclear
activity minor mergers might rather increase 
the existing AGN luminosity to QSO levels.
It is possible that the previously existing
Seyfert nucleus of I~Zw~1 is 
raised to QSO luminosities by the increased AGN fueling
rate due to the minor merger with the western companion
galaxy. The onset of the nuclear activity in 
I~Zw~1 might instead be related 
to a merger-independent AGN triggering mechanism.
A possible mechanism for merger-independent triggering
of Seyfert activity could be the 
stochastic accretion model
shown by \citet{2006ApJS..166....1H}.
As discussed above, the AGN triggering by a previous strong merger event
in the case of I~Zw~1 is unlikely because of
the clear spiral host galaxy.

\section{Summary\label{sec:summary}}

\begin{enumerate}

\item An ongoing tidal interaction between I~Zw~1
and the western source is likely to be responsible for
enhanced star formation activity in the western part
of the I~Zw~1 host galaxy.
This is suggested by the concentration of blue color
in the part of the I~Zw~1 host galaxy which is adjacent to
the western companion galaxy.

\item In agreement with previous studies, the 
NIR spectra and the optical-to-NIR 
color images of the western companion 
galaxy indicate an
old evolved stellar population without recent star
formation. This suggests that 
the western companion galaxy is a gas-poor dwarf
elliptical. An extension to the west of the likely
companion resolves into a separate object.

\item The NIR spectra of the northern source
reconfirm the scenario of a late-type foreground star,
as reported by \citet{1970ApJ...160..405S} and
\citet{1982ApJ...257...33S}.

\item In agreement with \citet{1998ApJ...500..147S},
the hydrogen line ratios indicate no significant amount
of reddening toward the very nucleus of I~Zw~1.
Out of the two 
blueshifted high-excitation lines reported by
\citet{1998ApJ...500..147S}
only $[\ion{Si}{vi}]$ is marginally detected.

\item The outskirts of the bulge region of I~Zw~1 
have redder colors than the disk. This might be
an indication for a predominance of old stars 
and/or an increased amount of dust in this region.
At the low angular resolution, the color might alternatively
be dominated by the NIR excess from the nucleus
\citep{1999ApJ...512..162S}. Higher angular 
resolution is needed for verifying the
previous signs for young stellar populations in the 
central region of the bulge
\citep{1998ApJ...500..147S, 2003A&A...405..959S}.

\end{enumerate}

\begin{acknowledgements}
We would like to thank the anonymous referee for 
a helpful report.
We are grateful to the anonymous 
observer who carried out the VLT observations in 
service mode. J. S. thanks Dr. M. Schirmer 
for his advice with the installation and use of the
THELI pipeline.
Parts of the research were supported
by the Deutsche Forschungsgemeinschaft (DFG)
via grant SFB~494 and by the ``Studienstiftung des deutschen Volkes''
via a scholarship for 
doctoral students for J. S. until June 2004.
\end{acknowledgements}

\bibliographystyle{aa}
\bibliography{6359.bib}

\end{document}